\documentclass[aps, prd, twocolumn, superscriptaddress, showkeys, byrevtex groupedaddress,nofootinbib]{revtex4-1}
\usepackage{scalerel}
\usepackage{tikz}
\usetikzlibrary{svg.path}

\definecolor{orcidlogocol}{HTML}{A6CE39}
\tikzset{
    orcidlogo/.pic={
        \fill[orcidlogocol] svg{M256,128c0,70.7-57.3,128-128,128C57.3,256,0,198.7,0,128C0,57.3,57.3,0,128,0C198.7,0,256,57.3,256,128z};
        \fill[white] svg{M86.3,186.2H70.9V79.1h15.4v48.4V186.2z}
        svg{M108.9,79.1h41.6c39.6,0,57,28.3,57,53.6c0,27.5-21.5,53.6-56.8,53.6h-41.8V79.1z M124.3,172.4h24.5c34.9,0,42.9-26.5,42.9-39.7c0-21.5-13.7-39.7-43.7-39.7h-23.7V172.4z}
        svg{M88.7,56.8c0,5.5-4.5,10.1-10.1,10.1c-5.6,0-10.1-4.6-10.1-10.1c0-5.6,4.5-10.1,10.1-10.1C84.2,46.7,88.7,51.3,88.7,56.8z};
    }
}

\newcommand\orcidicon[1]{\href{https://orcid.org/#1}{\mbox{\scalerel*{
                \begin{tikzpicture}[yscale=-1,transform shape]
                \pic{orcidlogo};
                \end{tikzpicture}
            }{|}}}}
\usepackage{dcolumn}  
\usepackage{bm}       
\usepackage{amssymb, amsmath,array}  
\usepackage{lipsum,graphicx,float}
\usepackage{subfig, caption}
\usepackage{url}
\usepackage{natbib}
\usepackage{mathtools}
\usepackage{txfonts}
\usepackage[normalem]{ulem}
\usepackage{booktabs}
\usepackage{epsfig}
\usepackage{hyperref}
\hypersetup{
    colorlinks=true,
    citecolor=blue,
    linkcolor=green,
    urlcolor=magenta}
\begin{document}
    \title{Constraint on Primordial Magnetic Fields In the Light of ARCADE 2 and EDGES Observations}

    \author{Pravin Kumar Natwariya \orcidicon{0000-0001-9072-8430}\,}
    \email{pravin@prl.res.in}
    \affiliation{%
        Physical Research Laboratory, Theoretical Physics Division, Ahmedabad 380 009, India}
    \affiliation{%
        Department of Physics, Indian Institute of Technology, Gandhinagar 382 424, India}  
    \date{\today}
\begin{abstract}
	\begin{center}
		{\bf Abstract}
	\end{center}
We study the constraints on primordial magnetic fields (PMFs) in the light of the Experiment to Detect the Global Epoch of Reionization Signature (EDGES) low-band observation and Absolute Radiometer for Cosmology, Astrophysics and Diffuse Emission (ARCADE 2).  ARCADE 2 observation detected extra-galactic excess radio radiation in the frequency range 3-90~GHz. The enhancement in the radio radiation is also supported by the first station of the Long Wavelength Array (LWA1) in the frequency range 40-80~MHz. The presence of early radiation excess over the cosmic microwave background can not be completely ruled out, and it may explain the  EDGES anomaly. In the presence of decaying PMFs, 21~cm differential brightness temperature can modify due to the heating of the gas by decaying magnetic fields, and we can constraint the magnetic fields. 

For excess radiation fraction ($A_r$) to be LWA1 limit, we show that the upper bound on the present-day magnetic field strength, $B_0$, on the scale of 1~Mpc is  $\lesssim 3.7$~nG for spectral index $n_B=-2.99$. While for $n_B=-1$, we get $B_0\lesssim1.1\times10^{-3}$~nG. We also discuss the effects of first stars on IGM gas evolution and the allowed value of $B_0$.  For $A_r$ to be LWA1 limit, we get the upper constraint on magnetic field to be $B_0(n_B=-2.99)\lesssim4.9\times10^{-1}$~nG and $B_0(n_B=-1)\lesssim3.7\times10^{-5}$~nG. By decreasing excess radiation fraction below the LWA1 limit, we get a more stringent bound on $B_0$.
\end{abstract}
\maketitle
\flushbottom
\clearpage
\section{Introduction}
\label{intro}
The 21~cm signal, due to the hyperfine transition between $1\rm S$ singlet and triplet states of the neutral hydrogen atom, is a  treasure trove to provide an insight into the period when the galaxies and first stars formed. Recently, the EDGES collaboration observed an absorption signal in the redshift range $15 \lesssim z \lesssim 20$. It is nearly two times more than the theoretical prediction based on the $\Lambda$CDM framework cosmological scenarios \cite{Bowman:2018yin, Pritchard_2012}. During the cosmic dawn, in the standard cosmological scenario, the temperature of the gas ($T_{\rm gas}$) and cosmic microwave background radiation (CMBR), $T_{\rm CMB}$, varies adiabatically. $T_{\rm gas}$ and $T_{\rm CMB}$ varies with the redshift  as $\propto(1+z)^2$ and $\propto(1+z)$ respectively, and temperatures of both the gas and CMBR found to be $\sim  6.7$~K and $\sim 49.1$~K at the redshift $z=17$ respectively (for example see the Ref. \cite{Seager1999, Seager, Chluba2015}). EDGES observation reported that the best fitting 21~cm model yields an absorption profile centred at $78\pm 1$~MHz and in symmetric ``U" shaped form having an amplitude of $T_{21}=-0.5_{-0.5}^{+0.2}$~K with 99\% confidence intervals \cite{Bowman:2018yin}. It is argued that to explain the EDGES observation, for the best fitting amplitude at the centre of the ``U" profile,  either the cosmic background radiation temperature $T_R\gtrsim104$~K for the standard  $T_{\rm gas}$ evolution or $T_{\rm gas}\lesssim3.2$~K in the absence of any non-standard evolution of the $T_R$, i.e. $T_R=T_{\rm CMB}$ \cite{Bowman:2018yin}. In the standard scenarios, background radiation is assumed to be solely contribution by the cosmic microwave background (CMB).


Although the contribution to the background radiation is assumed to be CMB radiation, the EDGES anomaly encouraged to develop the alternative models in which radio background enhanced \cite{Bowman:2018yin, Feng2018, Mondal:2020}. Recently, the Absolute Radiometer for Cosmology, Astrophysics and Diffuse Emission (ARCADE 2) collaboration, a double-nulled balloon-borne instrument with seven radiometers, detected the excess radio radiation in frequency range 3-10~GHz. It agrees with CMBR at the large frequency ($>$1~GHz) but significantly deviates at small frequency \cite{Fixsen2011, Feng2018}. This radio radiation is several times larger than the observed radio count due to the known processes \cite{Singal_2018}. Although in the Ref. \cite{Fang:2016}, authors discuss that merger of the clusters can generate the radio excess in the presence of magnetic turbulence,  the presence of early excess radiation can not be completely ruled out. For example, in the redshift range $z\approx30$ to 16, accretion onto the first intermediate-mass Black Holes can produce a radio radiation \citep{Ewall-Wice2018}. Subsequently, accreting supermassive black holes  \cite{Biermann:2014} or supernovae \cite{Jana:2018} can also produce radio background due to synchrotron emission at the time of cosmic down by accelerated electrons in the presence of the magnetic field.  The enhancement in the background radiation is also supported by the first station of the Long Wavelength Array (LWA1) in frequency range 40-80~MHz, and it is modelled by a power law with a spectral index ($\beta$) of $-2.58\pm0.05$ \cite{Dowell2018}, while ARCADE 2 is modelled with $\beta=-2.62\pm0.04$  \cite{Fixsen2011, Feng2018}.


Origin and evolution of primordial magnetic fields (PMFs) are one of the outstanding problems of cosmology (Ref. \cite{Subramanian:2016} and references therein). Presence of decaying PMFs can heat the gas above the $6.7$~K at $z=17$, and even it can erase the EDGES absorption signal \cite{Minoda:2018gxj, Sethi:2004pe, Chluba2015}. Still, the EDGES absorption signal can be explained by considering the possible early excess of radio radiation \cite{Feng2018}. In the present work, we consider decaying magnetohydrodynamics (MHD) and constraint the present-day strength of primordial magnetic fields. Observations suggest that the magnetic fields (MFs) are present on the length scale of galaxies to the clusters. Recently in the Ref. \cite{Jedamzik:2020L}, authors show that PMFs can be used as a remedy to resolve the Hubble tension between different observations. The present-day amplitude of these MFs is constrained from the big bang nucleosynthesis, formation of structures and cosmic microwave background anisotropies and polarization \cite{ Trivedi:2012ssp, Trivedi:2013wqa, Sethi:2004pe}. Authors of the Ref. \cite{Minoda:2018gxj}, put a upper constraint on PMFs strength $B_{\rm 1Mpc}\lesssim 10^{-10}$~G at the length-scale of 1~Mpc by considering $T_{\rm gas}\lesssim T_{\rm CMB}$ (i.e. $T_{21}\lesssim0$) so that, PMFs can not erase the absorption signal in the redshift range $15\lesssim z\lesssim20$. Planck 2015 results put upper constraints on PMFs of the order of the $\sim10^{-9}$~G for different cosmological scenarios \cite{Planck:2016}. The authors of the Ref. \cite{Natwariya:2020}, in the context of EDGES observation, put an upper and lower constraint on the PMFs to be $6\times 10^{-3}~{\rm nG} $ and $5\times 10^{-4}~{\rm nG}$ respectively.  Also, the lower bound on the present-day strength of PMFs found in Refs. \cite{Ellis:2019MMVA, Fermi_LAT:2018AB, Tavecchio:2010GFB}. Subsequently, in the Ref. \cite{Neronov:1900zz}, authors put a lower bound on the strength of intergalactic magnetic fields of the order of $3\times10^{-16}$~G using Fermi observations of TeV blazars. Upper constraint on the PMFs at the end of big bang nucleosynthesis found to be $2\times10^9$~G \cite{Cheng:1996vn}. Presence of strong PMFs can modify the present-day relic abundance of He$^4$ and other light elements. Therefore, Using the current observation of light element abundances, present-day MFs can be constrained \cite{Matese:1969cj, Greenstein:1969,Tashiro:2005ua, Sethi:2004pe, Choudhury:2015P}. The authors of the Ref. \cite{Jedamzik:2019}, put a constraint on the upper bound of PMFs strength of $47$~pG for scale-invariant PMFs by comparing CMB anisotropies, reported by the WMAP and Planck, with calculated  CMB  anisotropies. Generation of the magnetic fields in the early Universe for the various cosmological scenarios has been studied in the earlier literature (for example see Refs. \cite{Quashnock:1989sl, Grasso:2000wj, Subramanian:2010, Pandey:2020, Ellis:2019MMVA}). It is to be noted that decaying MHD has been studied in several literatures. In these works, the authors consider the decay of the PMFs by ambipolar diffusion and turbulent decay  \cite{Sethi:2004pe, Chluba2015, Bhatt2019pac, Minoda:2018gxj, Bera:2020}. Ambipolar diffusion of magnetic fields is important in neutral medium as it is inversely proportional to free-electron fraction ($X_e$) and $X_e\sim 10^{-4}$ after redshift $z\lesssim100$ \cite{Chluba2015, Peebles:1968ja, Sethi:2004pe}. Magnetic energy dissipation into gas, due to ambipolar diffusion, happens because of relative velocity between ion and neutral components of gas \cite{Shu:1992fh}. After the recombination ($z\sim1100$), the radiative viscosity of fluid dramatically decreases, and velocity perturbations are no longer damped, and tangled magnetic fields having length scale smaller than the magnetic Jeans length can dissipate via another mode--turbulent decay \cite{Sethi:2004pe, Chluba2015, Schleicher:2008aa}. Magnetic heating of the gas due to the turbulent decay decreases with redshift but later when ionization fraction decreases, heating increases due to ambipolar diffusion \cite{Chluba2015, Sethi:2004pe}.  


In the present work, we use the EDGES signal in the presence of excess radio radiation to constrain the strength of PMFs.  Some of the processes which we have discussed responsible for the excess radio radiation can occur at earlier redshift ($z\sim 17$) \cite{Ewall-Wice2018, Biermann:2014, Jana:2018}. One the interesting proposal in the Ref. \cite{Feng2018}, is to argue that such a possibility can exist at the earlier time also, and it can help to explain the EDGES signal. Authors show that the absorption signal can be explained by having excess radio radiation which is around 10\% of the observed (excess) radiation of ARCADE 2. In  Ref. \cite{Lawson:2019, Lawson:2013}, the authors claim that thermal emission from the axion quark nugget dark matter model can explain the EDGES signal, and it can also contribute a fraction of the radiation excess observed by ARCADE 2. At present, there exist several theoretical models to explain this excess at the time of cosmic down.  Recently it was argued that, stimulated emission from Bose (axion) stars can give a large contribution to the radio background possibly explaining EDGES and ARCADE 2 observations \cite{Levkov:2020}. In the Ref. \cite{Mebane:2020}, the authors consider accreting Pop III black holes and shows that radio emission from these sources can produce the EDGES like signal by increasing background radiation temperature. In other scenarios,  the  EDGES anomaly can be explained by axion-photon conversion in the presence of intergalactic magnetic fields \cite{Moroi2018} or by radiative decays of standard model neutrino induced by magnetic fields \cite{AristizabalSierra2018}. Radio excess can also be explained by the cusp region of superconducting cosmic strings \cite{Brandenberger:2019}. In ref. \cite{Chianese:2019}, authors consider radiative decays of relic neutrino and show that it can potentially explain the ARCADE 2 excess together with the EDGES signal. Depending on the origin, the excess fraction of radio radiation can have a different value. We discuss the constraints on excess radiation later.  Considering the above possibilities of having early excess radiation, we believe that it is important to analyze constraints on the primordial magnetic field in the presence of such radiation.


This work is organized in the following sections: In section \eqref{S_21cm}, we discuss the 21 cm signal due to the hyperfine transition between triplet and singlet state of the neutral hydrogen atom. We also discuss 21 cm differential brightness temperature due to the deviation of spin temperature from the background radiation temperature. In section \eqref{S_Tgas}, the evolution of the gas temperature and ionization fraction in the presence of decaying PMFs is discussed. Next, in section \eqref{Heat_result}, we consider the effects on the IGM temperature due to first stars. In section \eqref{S_result}, we discuss our results and obtain upper  constraint on the present day strength of PMFs in the absence/presence of x-ray and VDKZ18 heating \cite{Venumadhav:2018}.   


\section{21 cm differential brightness temperature}\label{S_21cm}

After the recombination, the baryon number density mostly dominated by the neutral hydrogen ($N_{\rm HI}$) and some fraction of residual free electrons ($X_e=N_e/N_H$) and protons ($X_p=N_p/N_H$). Here, $N_e$, $N_p$ and $N_H$ are number density of free electrons, protons and hydrogen nuclei respectively. The hyperfine interaction in neutral hydrogen atom splits it's ground state into $1\rm S$ triplet ($n_1$) and singlet ($n_0$) hyperfine levels. The Relative number density of hydrogen atom in triplet ($n_1$) and singlet ($n_0$) state is characterized by spin temperature ($T_S$), 
\begin{equation}
\frac{n_{1}}{n_{0}} = \frac{g_{1}}{g_{0}}\times \exp{(-{2\pi\nu_{10}}/{T_S})}\,, \label{eq1}
\end{equation}
here, $g_1$ and $g_0$ are statistical degeneracy of triplet and singlet states respectively and $\nu_{10}=1420$~MHz $=1/(21 ~ \text{cm})$ is corresponding frequency for hyperfine transition. In the context of cosmological scenarios, the spin temperature may depend on collisions between hydrogen atoms, absorption/ emission of background radiation and Ly-$\alpha$ radiation emitted from the first stars. Therefore, the spin temperature can be defined by requiring equilibrium balance between the populations of triplet and singlet state \cite{Field,Pritchard_2012, Furlanetto2006a},
\begin{alignat}{2}
T_S^{-1}=\frac{T_R^{-1} + x_\alpha T_\alpha^{-1}+x_cT_{\rm gas}^{-1} }{1+x_\alpha+x_c}\,\label{eq2}.
\end{alignat}
Here, $T_{\rm gas}$ is the kinetic temperature of the gas, and $T_R$ is the background radiation temperature.  As discussed in the introduction, the possibility of an excess radio radiation background over the CMBR can not be denied. For the excess radio background, we consider the phenomenological model following the Ref. \cite{Fialkov:2019}. Here, Authors consider a uniform redshift-independent synchrotron-like radiation, motivated by the ARCADE2 and LWA1 observations. This model can explain the EDGES anomaly in addition to enhancement of cosmic down power spectrum.  Accordingly, following the Refs. \cite{Fialkov:2019, Reis:2020, Yang:2018, Mondal:2020,banet:2020}, 
\begin{alignat}{2}
T_R=T_0\,(1+z)\,\left[1+A_r\,\left(\frac{\nu_{\rm obs}}{78~{\rm MHz}}\right)^\beta\ \right] \,,\label{eq3}
\end{alignat}
where, $T_0 =2.725$~K is the present day CMB temperature and $\beta=-2.6$ is the spectral index. $A_r$ is the amplitude defined relative to the CMB at reference frequency of 78~MHz. For the 21~cm signal $\nu_{\rm obs}$ is  $1420/(1+z)$~MHz. Authors of the Ref. \cite{Fialkov:2019}, put a limit on the excess radiation background to $1.9<A_r<418$ at reference frequency of 78~MHz by considering the effect of an uniform radiation excess on the 21~cm signal from the cosmic dawn, dark ages and reionization. Authors consider a synchrotron-like spectrum with spectral index $-2.6\,$. The case with $A_r\sim418$ corresponds to the LWA1 limit on $A_r$ at the reference frequency of 78~MHz \cite{Dowell2018, Fialkov:2019}. The stringent constraint on excess radiation comes from the Low-Frequency Array (LOFAR) to  $A_r<182$ (95 percent CL) and $A_r<259$ (99 percent CL)  at a reference frequency of 78~MHz and spectral index $-2.6\,$ \cite{Mondal:2020}. $T_{\alpha}\approx T_{\rm gas}$ is the colour temperature due to Ly$\alpha$ radiation from the first stars \cite{1952AJ.....57R..31W,Field}. $x_c$ and $x_\alpha$ are collisional and Wouthuysen-Field (WF) coupling coefficients, respectively \cite{1952AJ.....57R..31W, Field, Hirata2006, Mesinger:2011FS},
\begin{alignat}{2}
x_{c} = \frac{T_{10}}{T_R}\frac{C_{10} }{A_{10} }\ , \ \  x_{\alpha} = \frac{T_{10}}{T_R}\frac{P_{01} }{A_{10} }\ , \label{eq4}
\end{alignat}
here, $T_{10}=2 \,\pi\, \nu_{10}=5.9 \times 10^{-6}$~eV and $C_{10}=N_ik_{10}^{iH}$ is collision deexcitation rate. $i$ stands for hydrogen atom, electron and proton. $k_{10}^{iH}$ is the spin deexcitation specific rate coefficient due to collisions of species $i$ with hydrogen atom \cite{Pritchard_2012}. $P_{01}=4P_{\alpha}/27$ and $P_\alpha$ is scattering rate of Ly$\alpha$ radiation \cite{Pritchard_2012}. $A_{10}=2.86\times 10^{-15}$~sec$^{-1}$ is the Einstein coefficient for spontaneous emission from triplet to singlet state.\\
The 21 cm differential brightness temperature is given by \cite{Bowman:2018yin, Zaldarriaga:2004, Pritchard_2012}, 
\begin{alignat}{2}
T_{21}\approx 23x_{\rm HI}\left[\frac{0.15}{\Omega_{\rm m }h^2}\,\frac{1+z}{10}\right]^{1/2}\left(\frac{\Omega_{\rm b}h^2}{0.02}\right)  \left(1-\frac{T_R}{T_S}\right)~{\rm mK}\,,\label{eq5}
\end{alignat}
here, $x_{\rm HI}=N_{\rm HI}/N_H$ is the neutral hydrogen fraction. For this work, we consider the following values for the cosmological parameters: $\Omega_{\rm m }=0.31$, $\Omega_{\rm b}=0.048$, $h=0.68$, $\sigma_8=0.82$ and $n_s=0.97$ \cite{Planck:2018}. As $T_{21}\propto(T_S-T_R)$, there can be three scenarios. If $T_S=T_R$ then $T_{21}=0$ and there will not be any signal. For the case when $T_S> T_R$, emission spectra can be observed, and when  $T_S< T_R$, it leaves an imprint of absorption spectra. 21 cm signal evolution can be described as: after recombination ($z\sim1100$) to $z\sim 200$, gas and cosmic background radiation shares same temperature and maintain thermal equilibrium due to the Compton scattering. Therefore, $T_{21}=0$ and the signal is not observed. After $z\sim 200$ until $z\sim 40$, gas decouples from background radiation and temperature falls as $T_{\rm gas}\propto (1+z)^2$. It implies early absorption spectra of 21 cm signal. Nevertheless, this signal is not observed due to the poor sensitivity of radio antennas. The sensitivity falls dramatically below 50~MHz. After $z\sim 40$ to the formation of the first star, number density and temperature of the gas are very small, hence, $x_c\rightarrow 0$. Therefore, there is no signal \cite{Barkana:2018nd,Pritchard_2012}. After the first star formation, gas couples to the spin temperature due to Ly$\alpha$ radiation emitted from the first star by Wouthuysen-Field (WF) effect \cite{1952AJ.....57R..31W, 1959ApJ...129..536F}. Therefore, $x_\alpha\gg 1,x_c$ and absorption spectra can be seen. After $z\sim 15$, x-ray emitted from active galactic nuclei (AGN) starts to heat the gas and emission spectra can be seen \cite{Pritchard_2012}. 


\section{Evolution of the gas temperature in the presence of PMF{\scriptsize s}} \label{S_Tgas}

In the presence of decaying magnetohydrodynamics effects, the gas temperature can increase. It can even increase above the background radiation and can erase the 21 cm absorption signal  reported by EDGES \cite{Sethi:2004pe, Schleicher:2008aa, Chluba2015, Minoda:2018gxj}. Therefore, present-day PMFs strength can be constrained by the EDGES observation in the presence of excess radiation reported by ARCADE 2 and LWA1 \cite{Bowman:2018yin, Feng2018, Fixsen2011, Kogut:2011, Dowell2018, Fialkov:2019}. In the presence of turbulent decay and ambipolar diffusion, thermal evolution of the gas with the redshift can be written as \cite{Shu:1992fh, Sethi:2008eq, Schleicher:2008aa,Sethi:2004pe, Chluba2015},
\begin{alignat}{2}
\frac{dT_{\rm gas}}{dz}=2\,\frac{T_{\rm gas}}{1+z}&+\frac{\Gamma_c}{(1+z)\,H}(T_{\rm gas}-T_{\rm CMB})\nonumber\\
&-\frac{2}{3\,N_{\rm tot}(1+z)\,H}(\Gamma_{\rm turb}+\Gamma_{\rm ambi})\,,\label{eq6}
\end{alignat}
Here,  $N_{\rm tot}=N_H(1+f_{He}+X_e)$, $f_{He}=0.079$ and $T_{\rm CMB}=T_0\,(1+z)$ is the cosmic microwave background (CMB) temperature. $H\equiv H(z)$ is the Hubble parameter. At early times, $T_{\rm gas}$ remains in equilibrium with CMB temperature due to Compton scattering. However, the gas temperature will not be strongly affected by the comparatively small amount of energy in the non-thermal radio radiation. Therefore, $T_{\rm gas}$ and $T_\alpha$ can be assumed independent of the excess radiation \cite{Feng2018}. $\Gamma_C$ is the Compton scattering rate, defined as,
\begin{eqnarray}
\Gamma_{C}= \frac{8 \sigma_T \rho_\gamma N_e}{3\,m_e\,N_{\rm tot}}\,,\label{eq7}
\end{eqnarray}
here, $\rho_\gamma = a_r T_{\rm CMB}^4$, $a_r=7.57\times10^{-16}~{\rm J\,m^{-3}K^{-4}}$ is the radiation density constant, $\sigma_T$ is the Thomson scattering cross-section and $m_e$ is the mass of electron. Change in the electron fraction with redshift \cite{Seager,Seager1999,Bhatt2019pac,AliHaimoud:2010dx}, 
\begin{alignat}{2}
\frac{dX_e}{dz}  =  \frac{1}{H(1+z)}\ & \frac{\frac{3}{4}R_{Ly\alpha} + \frac{1}{4} \Lambda_{2s,1s}}{\beta_{B}+\frac{3}{4}R_{Ly\alpha}+\frac{1}{4}\Lambda_{2s,1s}}\nonumber \\
&\times\,\Big(N_H X_e^2\alpha_{B}-4(1-X_e)\,\beta_{B}e^{-E_{21}/T_{\rm CMB}} \Big)\,,
\label{eq8}
\end{alignat}
here, $\alpha_{B}$ is the case-B recombination coefficient and $\beta_{B}$ is the photo-ionization rate. $E_{21} = 2\pi/ \lambda_{\text{Ly}\alpha} $, $\lambda_{\text{Ly}\alpha}=121.5682\times10^{-9}$~meter is the hydrogen Ly$\alpha$ rest wavelength \cite{Seager1999}. $\Lambda_{2s,1s}=8.22~ {\rm sec}^{-1}$ is the two photon decay rate of hydrogen and $R_{\text{Ly}\alpha}=\frac{8 \pi H}{3 N_H (1-X_e)\lambda_{\text{Ly}\alpha}^3}$ is the Ly$\alpha$ photon escape rate \cite{AliHaimoud:2010dx}. Heating rate per unit volume due to the ambipolar diffusion ($\Gamma_{\rm ambi}$) and turbulence decay ($\Gamma_{\rm turb}$) is given by \cite{Sethi:2004pe, Chluba2015},
\begin{alignat}{2}
&\Gamma_{\rm ambi}=\frac{(1-X_e)}{\gamma\, X_e\, (M_HN_b)^2}\ \frac{|(\bm \nabla\times\bm B)\times\bm B|^2}{16\,\pi^2}\,,\label{eq9}\\
&\Gamma_{\rm turb}=\frac{1.5\ m\ \left[\ln(1+t_i/t_d)\right]^m}{\left[\ln(1+t_i/t_d)+1.5\ln\{(1+z_i)/(1+z)\}\right]^{m+1}}H\,E_B\,,\label{eq10}
\end{alignat}
here, $E_B=B^2/(8\pi)$ is the magnetic field energy density,
\begin{equation}
\frac{dE_B}{dz}=4\,\frac{E_B}{1+z}+\frac{1}{H\ (1+z)}\ (\,\Gamma_{\rm turb}+\Gamma_{\rm ambi}\,)\,,\label{11}
\end{equation}
and $m=2(n_B+3)/(n_B+5)$.  $z_i=1088$ is the redshift when heating starts due the magnetic fields (recombination epoch), $\gamma=1.9\times 10^{14}\,(T_{\rm gas}/{\rm K})^{0.375}{\rm cm}^3/{\rm g}/{\rm s}$ is the coupling coefficient, $M_{\rm H}$ is the mass of Hydrogen atom and $N_b$ is the number density of baryons. $t_d=1/\big(k_d\,V_A(k_d,z)\big)$ is the decay time for the turbulence.  For matter dominated era, $t_i=2/\big(3\,H(z_i)\big)$ and $V_A(k_d,z)=B(k_d,z)/\big(4\,\pi\,\rho_b(z)\,\big)^{1/2}\,$ is the Alfv\'{e}n wave velocity. $B(k_d,z)$ is the magnetic field strength smoothed over the scale of $k_d$ at redshift $z$. $k_d$ is constrained by the damping wavenumber of Alfv\'{e}n wave. PMFs having wavenumber larger than $k_d$, are strongly damped by the radiative-viscosity \cite{Sethi:2004pe, Schleicher:2008aa, Jedamzik1998, Kunze_2014, Subramanian:1997gi, Mack:2002}. Following the Ref. \cite{Minoda:2018gxj}, we take the time evolution of the Alfv\'{e}n wave damping scale. It is given as $k_d(z)=k_{d,\rm i}\,f(z)$ and $f(z_i)=1$. Here, $k_{d,\rm i}$ is the damping wavenumber at recombination epoch,
\begin{alignat}{2}
k_{d,\rm i}=2\pi~{\rm Mpc^{-1}}\, \Bigg[1.32\times &10^{-3} \left(\frac{B_0}{\rm nG}\right)^2\,\left(\frac{0.02}{\Omega_{\rm b}h^2}\right)\,\nonumber\\
&  \qquad\times\left(\frac{\Omega_{\rm m}h^2}{0.15}\right)^{1/2}\Bigg]^{-\frac{1}{n_B+5}}\label{kdi} \,.
\end{alignat} 
Here, to smooth the  magnetic field amplitude over the length scale of $k_{d,\rm i}\,$, we choose the Gaussian window function in Fourier space ($k$) as \cite{Caprini:2004,Minoda:2018gxj,Planck:2016},
\begin{alignat}{2}
B_{k_{d,\rm i}}^2=\int_{0}^{\infty}\frac{d^3k}{(2\pi)^3}\ {\rm e}^{-k^2\big(\frac{2\pi}{k_{d,\rm i}}\big)^2}\,P_B(k)=B_0^2\left[\frac{k_{d,\rm i}}{2\pi~{\rm Mpc^{-1}}}\right]^{n_B+3}\label{Bkd}\,.
\end{alignat}
Here, we consider PMFs power spectrum, $P_B(k)$, as power law in Fourier space  \cite{Minoda:2018gxj},
\begin{alignat}{2}
P_B(k)=\frac{(2\pi)^2}{\Gamma\big[(n_B+3)/2\big]}\ B_0^2\ \left(\frac{k}{\rm Mpc^{-1}}\right)^{n_B}~{\rm Mpc^3}\label{PB}\,.
\end{alignat} 
The magnetic field strength on the scale of 1~Mpc, $B_{1~\rm Mpc}^2=\int\, (dk/2\pi)^3\ \exp[-(k/{\rm Mpc^{-1}})^2]\ P_B(k)=B_0^2\,$. As discussed earlier, magnetic fields are strongly damped above inverse length-scale ($k_d$), therefore, $P_B(k)=0$ for $k\ge k_d$. Lorentz force and the magnetic energy density can be calculated as \cite{Minoda:2018gxj},
{\small\begin{equation}
    |(\bm \nabla\times\bm B)\times\bm B|^2=\int_{k_1,k_2}k_1^2\ P_B(k_1)\ P_B(k_2)\ f^{2n_B+8}(z)\ (1+z)^{10}\,,\label{LF}
    \end{equation}}
here $\int_{k_1,k_2}[\cdots]=\int \int d^3k_1/(2\pi)^3 \times d^3k_2/(2\pi)^3\,[\cdots]$, and
\begin{equation}
E_B=\frac{1}{8\pi}\,\int\frac{d^3k}{(2\pi)^3}\  P_B(k)\ f^{n_B+3}(z)\ (1+z)^{4}\,.\label{EB}
\end{equation}
We can get the redshift evolution of the function $f(z)$, by substituting equation \eqref{EB} in equation \eqref{11}.


\section{Heating of the IGM due to background radiation}\label{Heat_result} 
\begin{figure*}
    \centering
    \subfloat[] {\includegraphics[width=2.2in,height=1.5in]{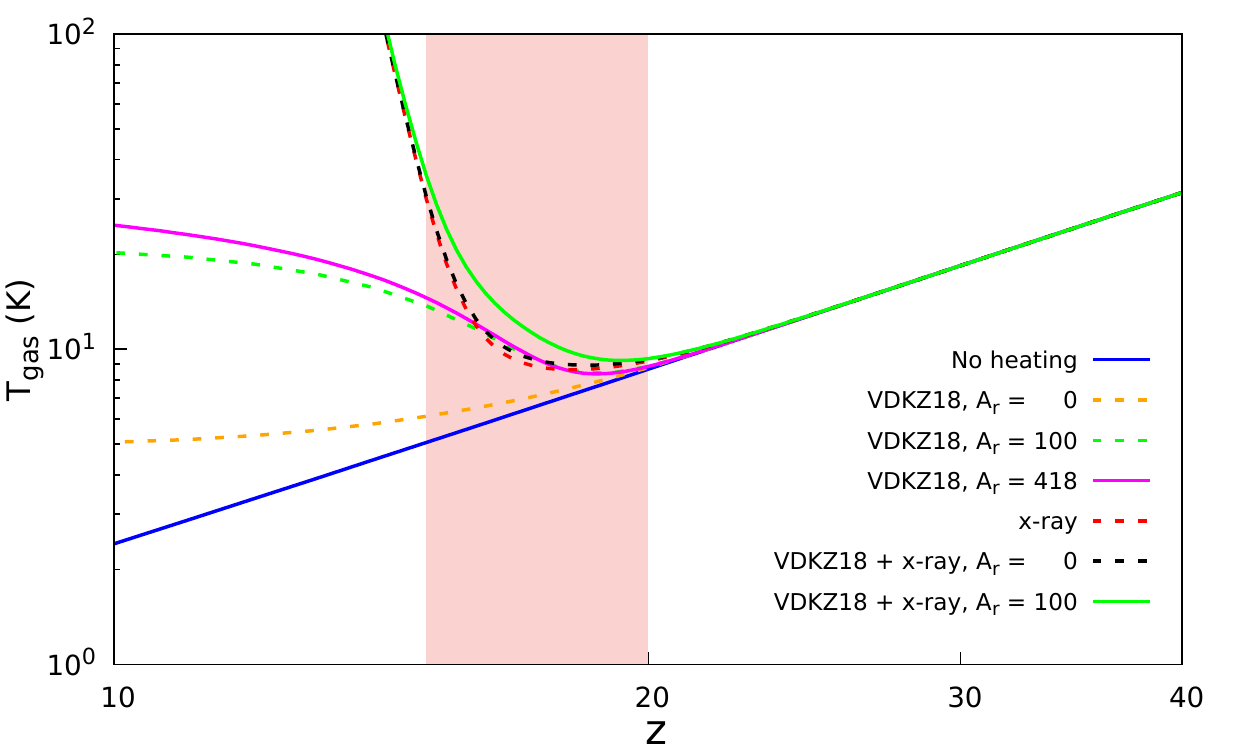}\label{p_1a}}
    \subfloat[] {\includegraphics[width=2.2in,height=1.5in]{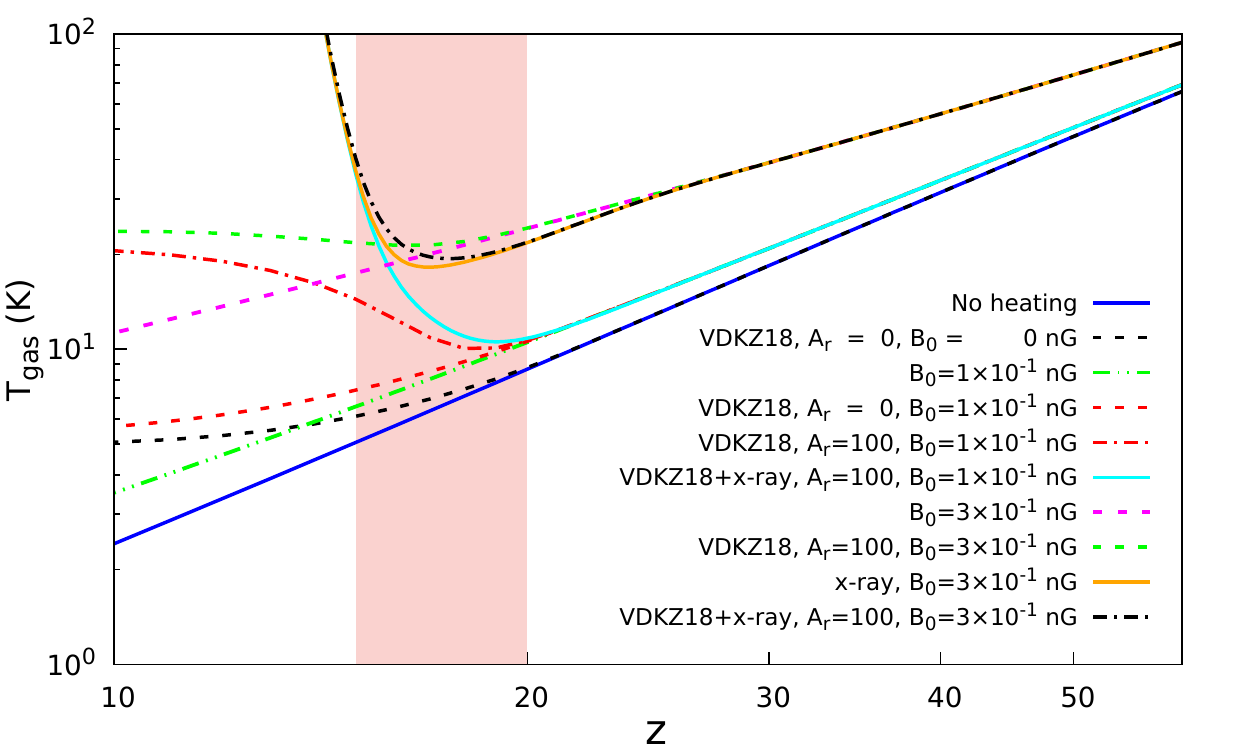}\label{p_1b}}
    \subfloat[] {\includegraphics[width=2.2in,height=1.5in]{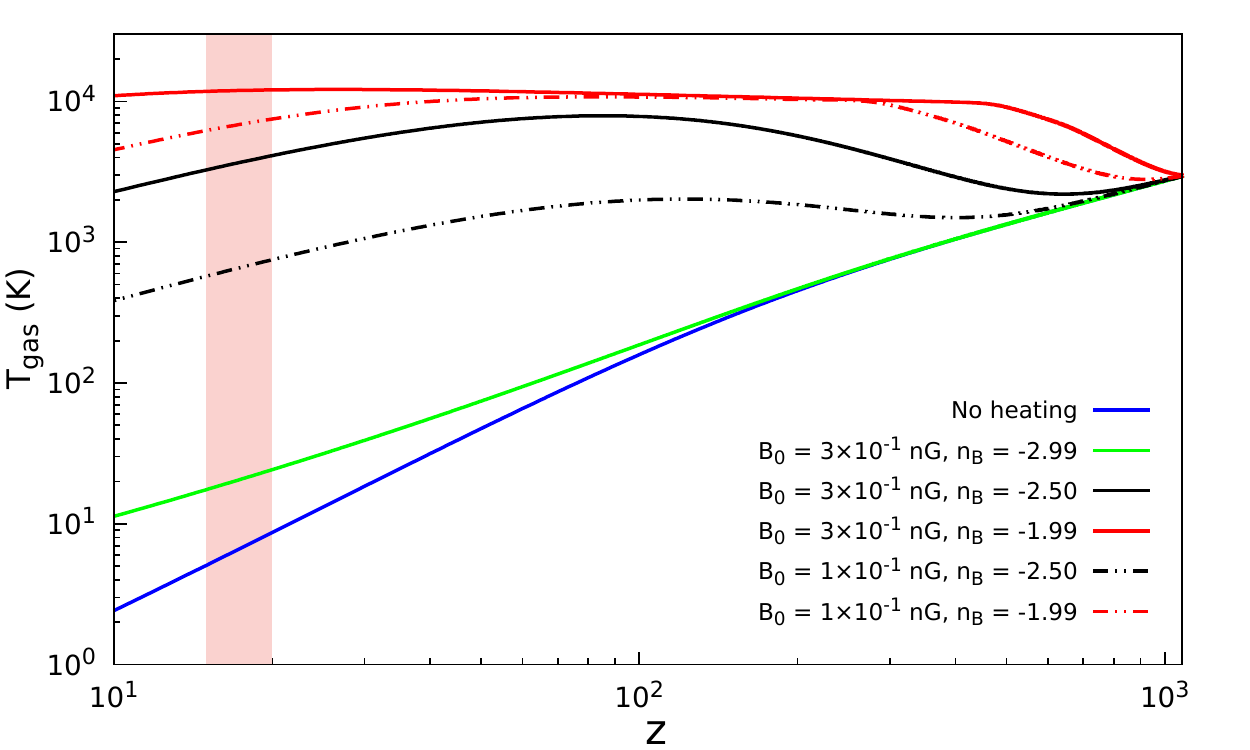}\label{p_1c}} 
    \caption{\raggedright The gas temperature evolution with redshift. The solid blue lines, in all plots, represent the case when there is no x-ray, VDKZ18 or magnetic heating. VDKZ18 corresponds to the heat transfer from the background radiation to gas mediated by Ly$\alpha$. The shaded region represents the EDGES observation redshift range, $15\leq z \leq 20$ . In figure \eqref{p_1a}, we consider only VDKZ18 and x-ray heating with excess radiation ($A_r$). In figure \eqref{p_1b}, we include different combination of VDKZ18, x-ray and magnetic heating, and spectral index is fixed to -2.99. In figure \eqref{p_1c}, we vary the spectral index and plot magnetic heating of the gas.}
    \label{p_1}
\end{figure*}

\begin{figure*}
    \centering
    \subfloat[] {\includegraphics[width=3.3in,height=2in]{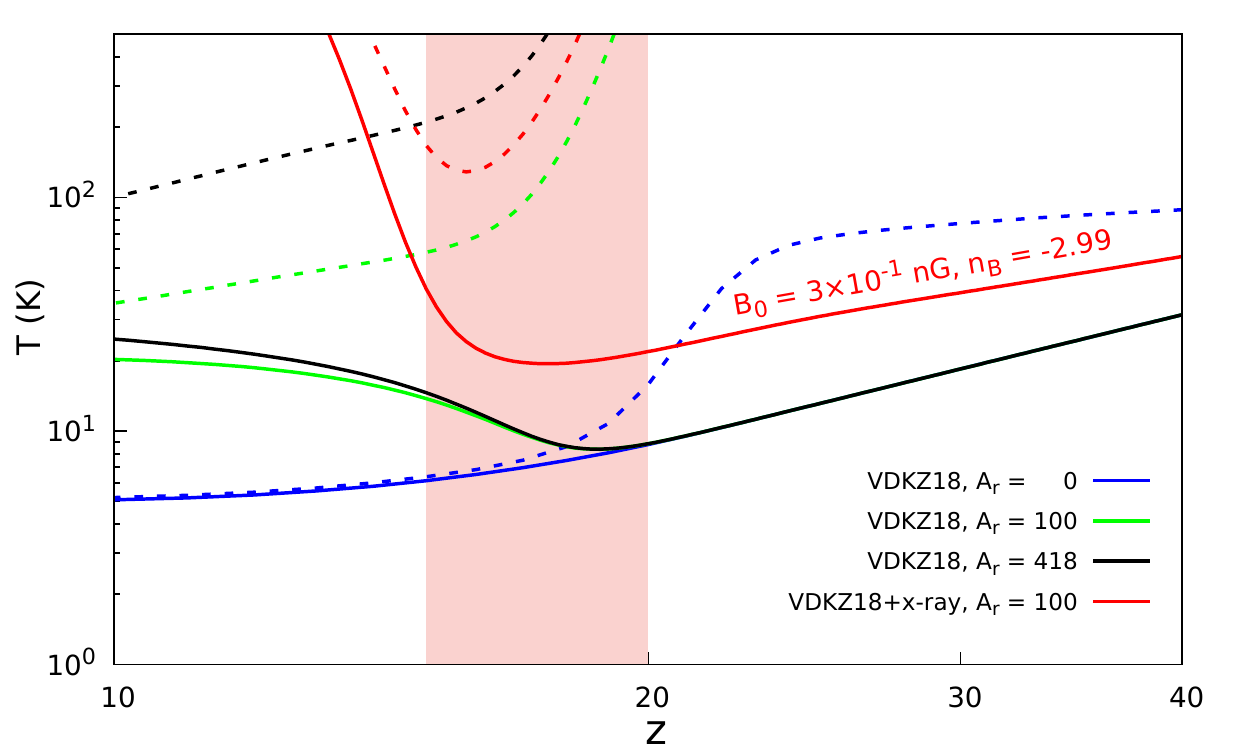}\label{p_2a}}
    \subfloat[] {\includegraphics[width=3.4in,height=2in]{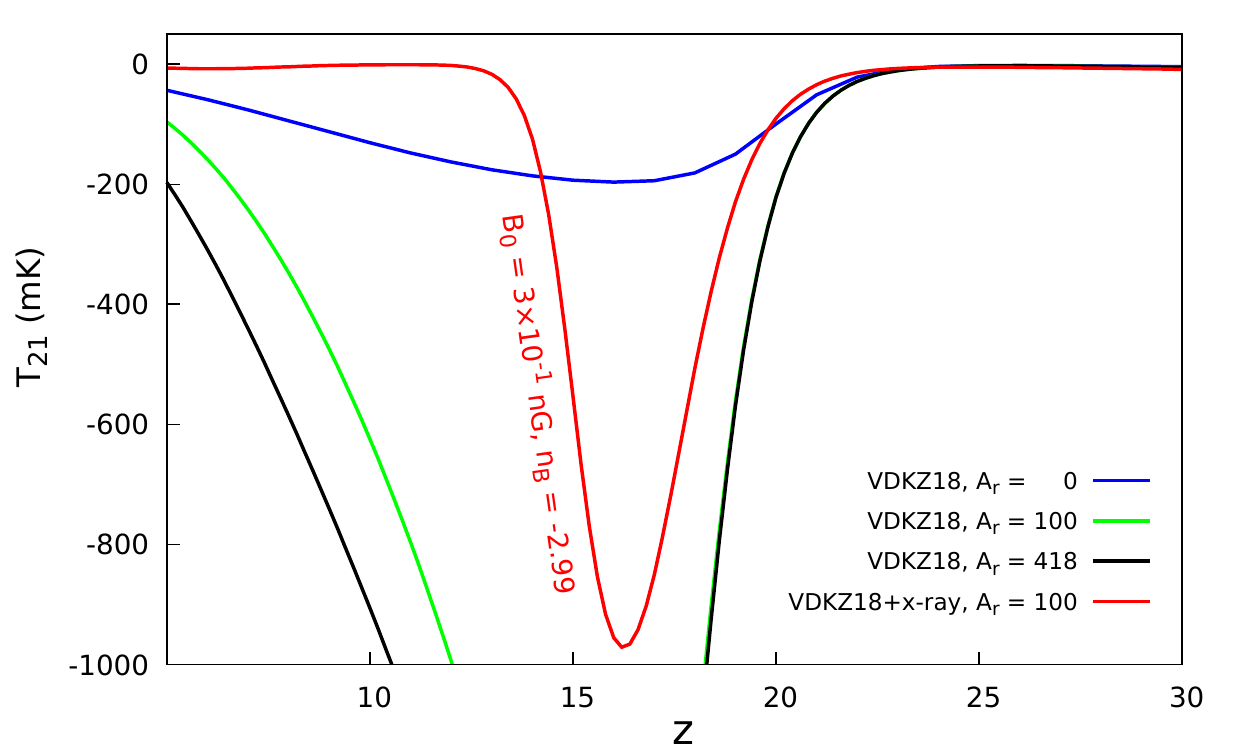}\label{p_2b}} 
    \caption{\raggedright Plot \eqref{p_2a} shows the gas (solid lines) and spin (dashed lines) temperature evolution, The shaded region corresponds to the redshift $15\leq z \leq 20$ ---the redshift range for EDGES reported signal. Plot \eqref{p_2b}, shows the 21 cm differential brightness temperature with redshift for same cases in plot \eqref{p_2a}.}
    \label{p_2}
\end{figure*}

After the first star formation ($z\sim30$), their radiation starts to heat the intergalactic medium (IGM) \cite{Furlanetto2006a, Ghara2019, Mirocha2019, Mesinger:2011FS, Mesinger2013a, Fialkov:2016A, Park:2019}. Authors of the Ref. \cite{Venumadhav:2018}, suggests that the kinetic temperature of the gas can also increase due the background radiation even in the absence of x-ray heating. The Ly$\alpha$ photons, due to first stars, intermediate the energy transfer between the thermal motions of the hydrogen and background radiation. Authors claim that this correction to the kinetic temperature of the gas is the order of ($\sim 10\%$) at $z=17$, in the absence of x-ray heating (hereafter we use the term VDKZ18 for this heating of the gas). Following the above reference, the equation \eqref{eq6}  will modify,
\begin{alignat}{2}
\frac{dT_{\rm gas}}{dz}=\frac{dT_{\rm gas}}{dz}\Bigg|_{[{\rm eq. \eqref{eq6}}]}+\frac{dT_{\rm gas}}{dz}\Bigg|_{\rm x-ray}-\frac{\Gamma_{R}}{(1+z)\,(1+f_{He}+X_e)}\,,\label{eq12}
\end{alignat}
where, ${dT_{\rm gas}}/{dz}\big|_{[{\rm eq. \eqref{eq6}}]}$ stands for the gas temperature evolution represented in equation \eqref{eq6}, and
\begin{alignat}{2}
\Gamma_{R}=x_{\rm HI}\,\frac{A_{10}}{2\, H}\,x_{R} \left[\frac{T_R}{T_S}-1\right]\,T_{10}\,,\label{eq13}
\end{alignat}
here, $x_R=1/\tau_{21}\times[1-\exp(-\tau_{21})]$, and the 21~cm optical depth $\tau_{21}=8.1\times10^{-2}\,x_{\rm HI}\,[(1+z)/20]^{1.5}\,(10~{\rm K}/T_S)$. And, $T_{10}=2\pi\nu_{10}=0.0682$~K. To include the x-ray heating of the IGM, we consider the $tanh$ parameterization \cite{Kovetz2018,Mirocha:2015G,Harker:2015M}. In the presence of x-ray radiation, the ionization fraction evolution with redshift will also change.  For the present case, we consider the fiducial model, for x-ray heating and ionization fraction evolution, motivated by Ref.  \cite{Kovetz2018}. The heating effects of both the VDKZ18 and x-ray are discussed in plots \eqref{p_1a}, \eqref{p_1b},  \eqref{p_2}, \eqref{p_3b} and \eqref{p_4b}.

%


\section{Result and discussion}\label{S_result}
\begin{figure*}
    \centering
    \subfloat[] {\includegraphics[width=3.4in,height=2in]{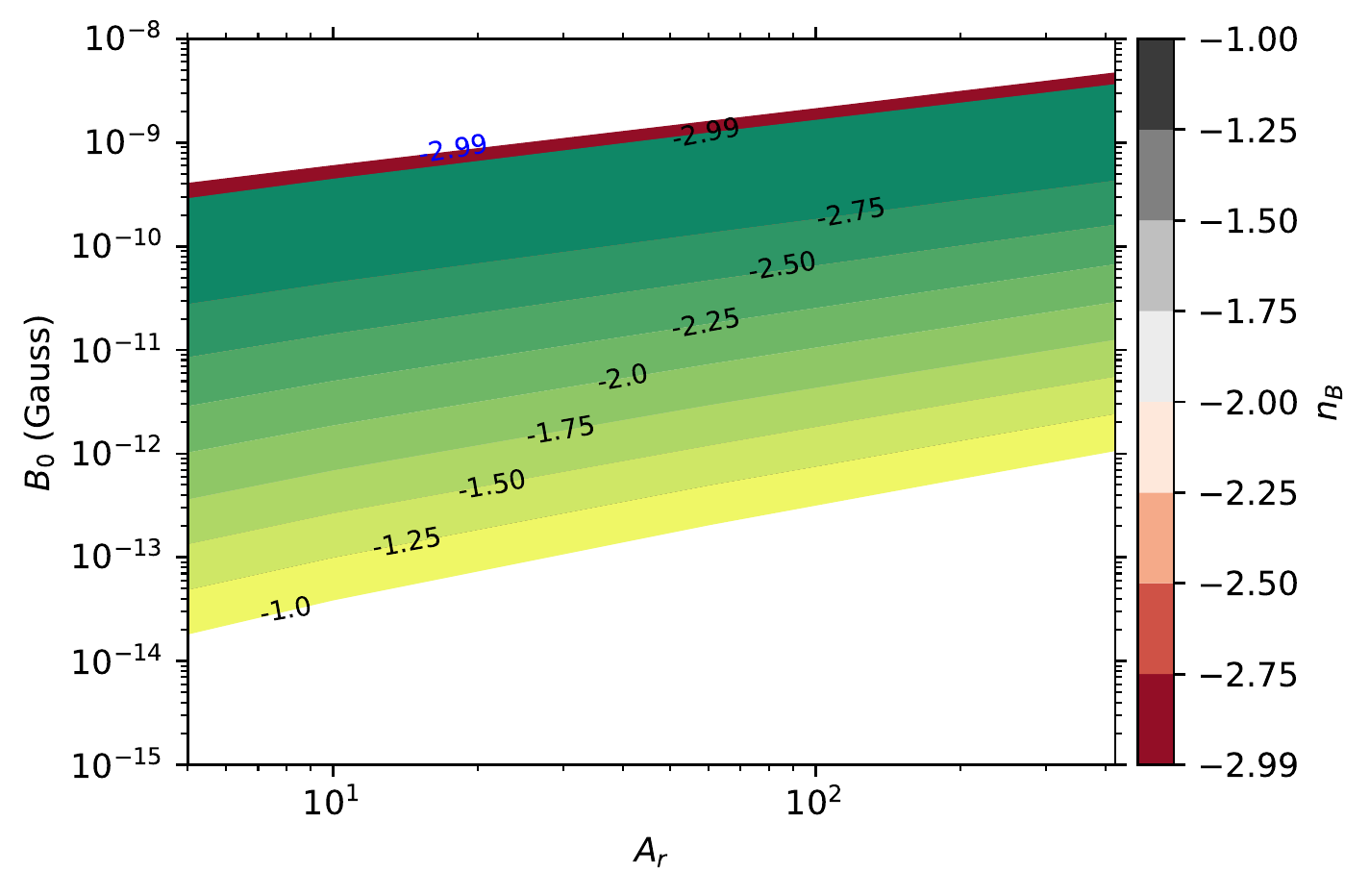}\label{p_3a}}
    \subfloat[] {\includegraphics[width=3.4in,height=2in]{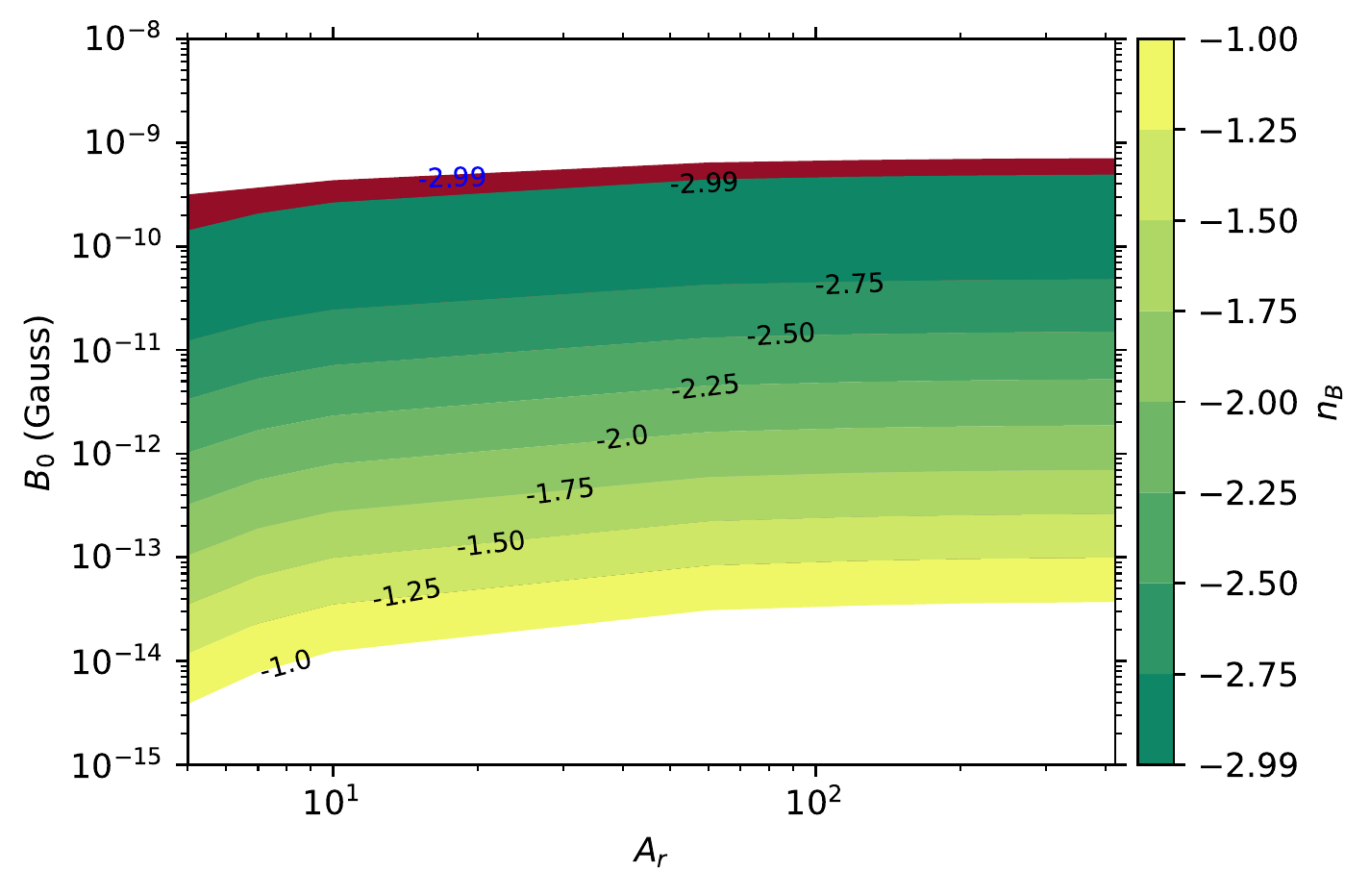}\label{p_3b}} 
    \caption{\raggedright In these plots, we study upper bounds on present-day magnetic field strength ($B_0$) with excess radiation fraction $(A_r)$ for different values of the spectral index, $n_B$. The green-yellow and red-grey colour schemes represent the cases when $T_{21}|_{z=17.2}\simeq-500$~mK and $-300$~mK, respectively. For $T_{21}|_{z=17.2}\simeq-300$~mK case the value of $n_B$ written with blue coloured text , while for $-500$~mK  case it is written with black coloured text. In plot \eqref{p_3a}, we consider $T_{\rm S}\simeq T_{\rm gas}$ and do not take into account the x-ray and VDKZ18 effects. While in figure \eqref{p_3b}, we consider the effects of VDKZ18 and x-ray on IGM gas due to first stars after $z\lesssim 35$ and consider finite Ly$\alpha$ coupling. The colour-bars are common for both plots.}
    \label{p_3}
\end{figure*} 
To study the gas temperature evolution with redshift in the presence of primordial magnetic field dissipation, we solve the coupled equations \eqref{eq6}, \eqref{eq8} and \eqref{11}. To get the Lorentz force term in equation \eqref{eq9}, we solve the equation  \eqref{LF}. Similarly, to get the magnetic field energy density in equation \eqref{eq10}, we solve the equation \eqref{EB}. To get the evolution of the $f(z)$ with redshift, $df(z)/dz$, we substitute equation \eqref{EB} in equation \eqref{11} with initial condition $f(z_i)=1\,$. To obtain upper constraint on PMFs strength, we solve the equation \eqref{eq5} with equations \eqref{eq6}, \eqref{eq8} and \eqref{11} for $T_{21}\simeq-300$~mK or -500~mK by varying $B_0$, $n_B$ and $A_r$. For infinite Ly$\alpha$ coupling $T_S\simeq T_{\rm gas}$, therefore, $T_S$ solely depends on the gas temperature. While, for finite Ly$\alpha$ coupling, $T_S$ depends on both the gas and background radiation temperature.   


In figure \eqref{p_1}, we plot the gas temperature evolution with the redshift for different present-day magnetic field strength and background radiation. The solid blue lines represent the case when there is no heating of the IGM gas, i.e. no x-ray, VDKZ18 or magnetic heating. The pink shaded band in the figure shows the EDGES redshift range, $15\leq z \leq 20$, for the 21~cm absorption signal. In plot \eqref{p_1a}, we consider only VDKZ18 and x-ray heating. The orange dashed line describes the heating due to  VDKZ18 only while keeping $A_r=0$.  Next, we increase the value of $A_r$ from 0 to 100.  This case is described by the dashed-green line in plot \eqref{p_1a}, which shows a significant rise in the gas temperature due to the excess radiation fraction. Further, if one increases the $A_r$  to its LWA1 limit, i.e. $A_r= 418$, the gas temperature does not change significantly from $A_r =100$ case, as shown by the solid magenta curve. It happens because $\Gamma_{R}\propto (T_R/T_S-1)\sim T_R/T_S$, equation \eqref{eq13}. As we increase $A_r$, $T_R/T_S$ increases slowly. For example, at $z=17$, $T_R/T_S$ is $6.5$ for $A_r=0$, $51.4$ for $A_r=100$ and $54.9$ for $A_r=418$. Here, we can see that, even increasing $A_r$ to $\sim4$ times (100 to 418), $T_R/T_S$ increases by only $6.8$ percent.   Therefore, increasing further $A_r$ will not affect gas temperature significantly. To analyse the role of x-ray heating, we have first considered the heating due to x-ray only, depicted by the red dashed line. The inclusion of VDKZ18 for $A_r=0$ further increases the gas temperature slightly, as shown by the black dashed line. In this case of inclusion of x-ray heating, if we increase the value of $A_r$ to 100, there is a significant increase in the gas temperature as shown by the solid green line. We find the contribution due to x-ray heating dominates for redshift values $z\lesssim15$. 


In plot  \eqref{p_1a}, we compare the contribution of VDKZ18 and x-ray heating. In plot \eqref{p_1b}, we compare the contributions of VDKZ18, x-ray and magnetic heating while keeping the spectral-index, $n_B=-2.99$ for a nearly scale-invariant magnetic field spectrum.  While in figure \eqref{p_1c}, we vary the magnetic spectral index ($n_B$) and plot the magnetic heating of the gas.


In plot \eqref{p_1b}, we have included the effect of primordial magnetic fields on the IGM gas evolution.  The solid blue line represents the case when there is no heating, and the dashed-black curve shows the case of VDKZ18 with no magnetic fields and x-ray for $A_r=0$. The double dot-dashed green curve represents the case when there is only the magnetic heating with a magnetic field strength of $B_0=1\times10^{-1}$nG.  Next, we include the case of VDKZ18 for $A_r=0$ in the pure magnetic heating scenario, as shown by the red dashed curve. Now, if we increase  $A_r$ from 0 to 100, the gas temperature rises significantly in the shaded region as shown by the dash-dotted red curve in figure \eqref{p_1b}. Now the further addition of x-ray heating is shown by the cyan plot, which shows significant heating in the shaded region.  Next, for more analysis, we increase the magnetic field strength from $B_0=1\times10^{-1}$~nG to $B_0=3\times10^{-1}$~nG and study cases with VDKZ18 and x-ray as before. The magenta dashed line depicts the case with only magnetic heating. The green dashed line shows the case of VDKZ18 with $A_r=100$. The orange curve shows the case with magnetic and x-ray heating only. Here, as expected, the gas temperature decreases after the inclusion of the x-ray effect with the magnetic fields. It happens because the ionization fraction increases by x-ray radiation. Ambipolar diffusion evolves as $\Gamma_{\rm ambi}\propto(1-X_e)/X_e$; therefore, as ionization fraction increases, ambipolar diffusion of the magnetic field decreases. Thus, the heating due to magnetic fields also decreases.  Therefore, including the x-ray contribution with the magnetic field decreases the magnetic field diffusion. Hence, the gas temperature decreases (this effect also occurs for $B_0=1\times10^{-1}$~nG, but it is not visible in the plot). The black dot-dashed line includes all the three effects: magnetic and x-ray heating together with VDKZ18 for $A_r=100$ and $B_0=3\times10^{-1}$~nG. Here, the addition of the VDKZ18 heating for $A_r=100$ increases the gas temperature above the solid orange line. It is also lower than the magenta dashed line because of the inclusion of the x-ray contribution.  At the smaller redshift, x-ray heating dominates over all other heating mechanisms, and all lines merge. 


In figure \eqref{p_1c}, we plot the magnetic heating of the gas for the different spectral index, $n_B$.  The solid lines, except the blue one, represent the magnetic heating for $B_0=3\times10^{-1}$~nG, while double dot-dashed lines are for $B_0=1\times10^{-1}$~nG. Increasing the spectral index, the magnetic heating due to ambipolar diffusion and turbulent decay increases as $\Gamma_{\rm ambi}\propto \big(1/\Gamma[(n_B+3)/2]\big)^2$ and  $\Gamma_{\rm turb}\propto 1/\Gamma[(n_B+3)/2]$ (by ignoring the logarithmic and integral dependencies). For example, if one changes $n_B$ from its value -2.99 to  $-1$ then $1/\Gamma[(n_B+3)/2]$ changes from $5\times10^{-3}$ to  1. Therefore, by increasing $n_B$ from -2.99 to -1,  magnetic heating enhances significantly. To get $T_{21}$ (equation \eqref{eq5}) around -500~mK or -300~mK at $z=17.2$, one needs to ensure that even by increasing  $n_B$, that the factor $x_{\rm HI}\left(1-{T_R}/{T_S}\right)$  remains same. Thus from equations \eqref{eq9}, \eqref{eq10} and \eqref{PB} when we increase $n_B$, we have to decrease $B_0$ so that the magnetic heating contribution to the gas remains the same. Therefore, by increasing $n_B$,  the upper bound on $B_0$ will become more stringent. Here, we also include the collisional ionization of the gas in equation \eqref{eq8}, as this term is important only when gas temperature is $\gtrsim1.58\times10^5$~K. Otherwise this term is exponentially suppressed as $\propto \exp[-(13.6~{\rm eV})/T_{\rm gas}]$     \cite{Sethi:2004pe, Asselin:1988, Shiraishi:2014}. In plot \eqref{p_1c}, the gas temperature rises by increasing $B_0$, as more magnetic energy is getting injected into thermal energy of the gas via $\Gamma_{\rm ambi}\propto E_B^2$ and $\Gamma_{\rm turb}\propto E_B$. However, for redshift $z\lesssim 100$, the gas temperature starts decreasing as the cooling effect due to expansion of the Universe become dominant, as can be seen in equations \eqref{eq6} \& \eqref{11} (it also depends on the strength and spectral index of the magnetic field). Since,  with the expansion of the Universe, magnetic energy density ($E_B$) also dilutes, the contributions from $\Gamma_{\rm ambi}$ and $\Gamma_{\rm turb}$ decreases as can be seen from equations \eqref{eq9}, \eqref{eq10} and \eqref{11}.


In figure \eqref{p_2a}, we plot the spin (dashed lines) and gas (solid lines) temperature.  For $A_r=0$, i.e. $T_R=T_{\rm CMB}$,  we get $T_{\rm gas}\simeq T_S$ as seen by the overlapping  dashed and solid  blue lines in the shaded region. $x_\alpha$ and $x_c$ are $\propto 1/T_R\,$ as can be seen from equations \eqref{eq2} and \eqref{eq4}. Therefore, the coupling between the gas and spin temperature decreases by increasing $A_r$. As discussed before, increasing the value of $A_r$ above $\sim 100$, the spin temperature increases, but the increment in gas temperature becomes insignificant, and the $T_R/T_S$ ratio increases slowly. Therefore, as $x_\alpha$ and $x_c$ decreases, the difference between the gas and spin temperature increases, as shown in the plot \eqref{p_2a}. Increasing the values of $A_r$ from $100$ (green lines) to $418$ (black lines), the difference between gas and spin temperatures increases. Figure \eqref{p_2b}, shows the plots for 21 cm differential brightness temperature vs. redshift,  for all the cases discussed in plot \eqref{p_2a}. As we increase the $A_r$ from 0 to 100 the $|T_{21}|$ increases. By increasing $A_r$  from 100 to 418, values of $T_{21}$ does not change significantly. Further, including x-ray heating and magnetic heating (for $B_0=3\times10^{-1}$~nG and $n_B=-2.99$) the gas temperature rises and $|T_{21}|$ decreases.


\begin{figure*}
    \centering
    \subfloat[] {\includegraphics[width=3.4in,height=2.1in]{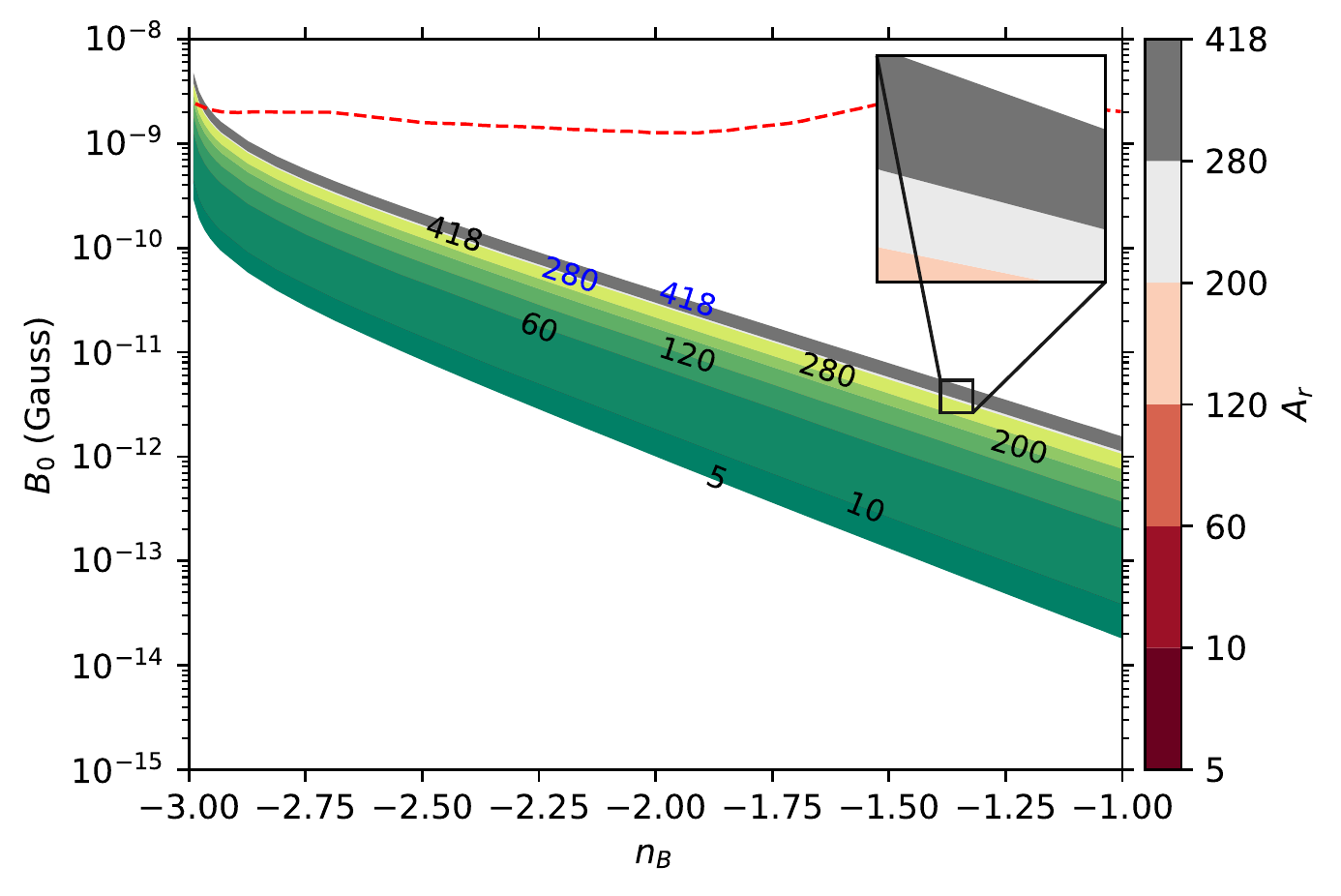}\label{p_4a}}
    \subfloat[] {\includegraphics[width=3.4in,height=2.1in]{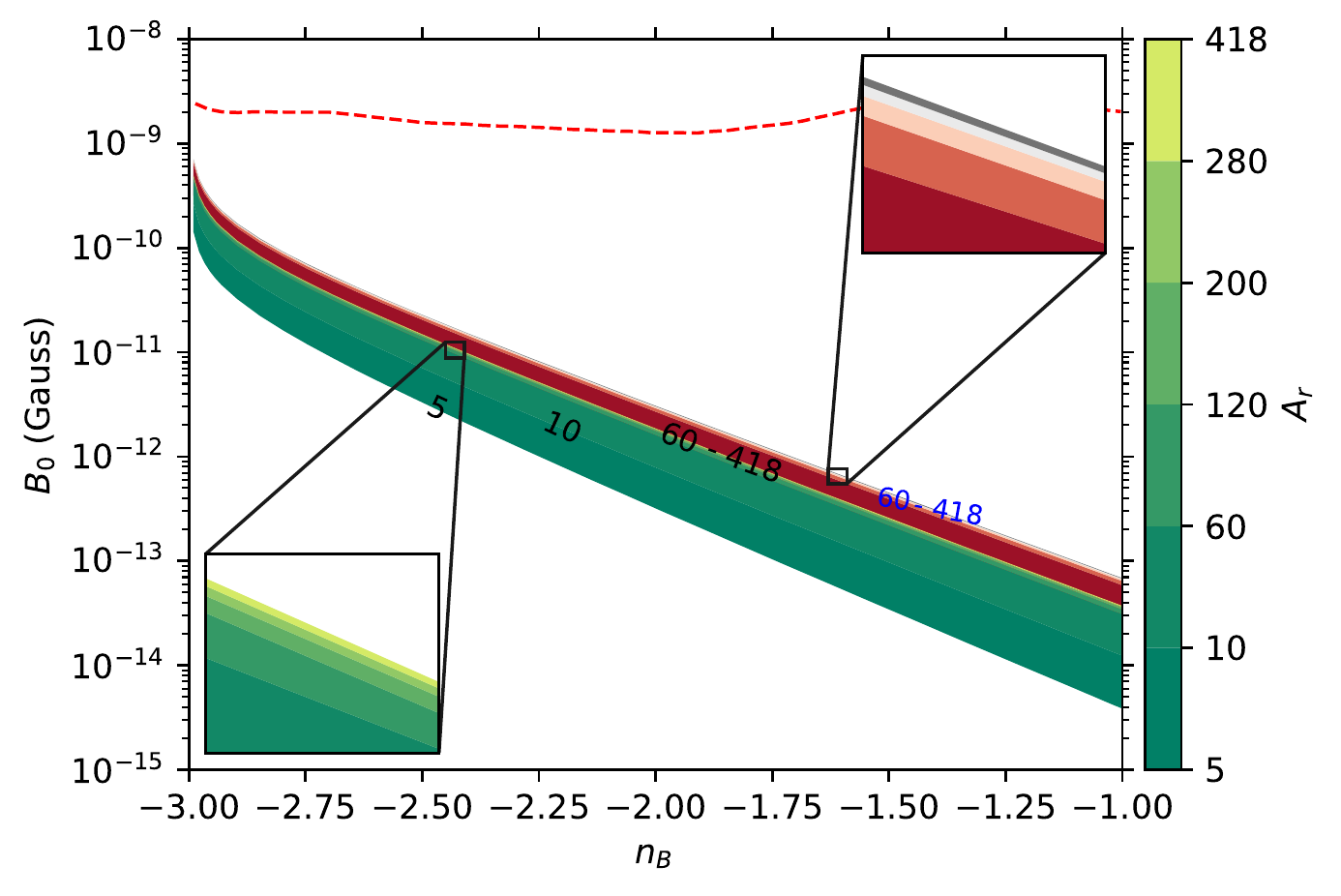}\label{p_4b}} 
    \caption{\raggedright In these plots, we study upper bounds on the present-day magnetic field strength ($B_0$) with spectral index ($n_B$) for different values of excess radiation fraction $(A_r)$. The green-yellow and  red-grey colour schemes represent the cases when $T_{21}|_{z=17.2}\simeq-500$~mK and $-300$~mK, respectively. For $T_{21}|_{z=17.2}\simeq-300$~mK case the value of $n_B$ written with blue coloured text , while for $-500$~mK  case it is written with black coloured text. In plot \eqref{p_4a}, we consider $T_{\rm S}\simeq T_{\rm gas}$ and do not take into account the x-ray and VDKZ18 effects. While in figure \eqref{p_4b}, we consider the effects of VDKZ18 and x-ray on IGM gas due to first stars after $z\lesssim 35$ and consider finite Ly$\alpha$ coupling. The colour-bars are common for both plots. The red dashed line depicts the Planck 2015 upper constraint on the present-day magnetic field strength \cite{Planck:2016, Minoda:2018gxj}.}
    \label{p_4}
\end{figure*} 


In figure \eqref{p_3}, we plot the maximally allowed values of $B_0$ versus radiation excess ($A_r$) for different spectral indexes. The colour-bars represent the variation of the magnetic field spectral index. In the plots, the spectral index varies from its nearly scale-invariant value (-2.99) to -1. Here, we consider both the EDGES best fit and upper constraint on the 21 cm absorption signal for constraining $B_0$. The green-yellow colour scheme represents the case with  $T_{21}|_{z=17.2}$ $\simeq-500$~mK, while the red-grey colour scheme represents the case with $T_{21}|_{z=17.2}\simeq-300$~mK. Numerical values of $n_B$ for the different colour bands are written with different colour. For $T_{21}|_{z=17.2}\simeq-300$~mK case the value of $n_B$ written with blue coloured text , while for $T_{21}|_{z=17.2}\simeq-500$~mK  case it is written with black coloured text. The colour-bars are common for both the plots.


In figure \eqref{p_3a}, we consider infinite Ly$\alpha$ coupling ($x_\alpha\gg x_c,\,1$), i.e. $T_S\simeq T_{\rm gas}$.  Here, we do not consider the x-ray and VDKZ18  effects on the  gas and thus the 21 cm signal $T_{21}\propto(1-T_R/T_{\rm gas})$. As we increase $A_r$, the amplitude of $|T_{21}|$ increases, and we get more window to increase the gas temperature. In this plot, we consider heating only due to the decaying magnetohydrodynamics. Therefore, we can increase $B_0$ as we increase $A_r$. As discussed earlier, by decreasing $n_B$, the amplitude of the magnetic field power spectrum also decreases, resulting in less magnetic energy dissipation into the gas kinetic energy. Thus by reducing values of $n_B$ from -1 to -2.99, we get more window to increase $B_0$. Next, when one increases  $T_{21}$ from -500~mK to -300~mK,  the allowed value of $B_0$ also increases. This is shown by the red-grey colour scheme in figure \eqref{p_3}.  In figure \eqref{p_3b}, we consider the effects of VDKZ18 and x-ray on IGM gas evolution due to first stars after $z\lesssim 35$ and consider finite Ly$\alpha$ coupling. As discussed earlier, $T_{\rm gas}\neq T_S$ for $A_r>0$ and the difference between gas and spin temperature increases as $A_r$ increases. Thus, in the presence of first star's effects, the upper bound on the present-day strength of PMFs modifies. Following the Refs. \cite{Kovetz2018, Mirocha:2015G, Harker:2015M}, we consider WF coupling coefficient, $x_\alpha = 2A_\alpha(z) \times (T_0/T_R)$. Here, $A_\alpha(z) = A_\alpha (1 + \tanh[(z_{\alpha0}-z)/\Delta z_\alpha])$, the step height $A_\alpha=100$, pivot redshift $z_{\alpha0}=17$ and duration $\Delta z_\alpha=2$. The collisional coupling coefficient, $x_{c} = T_{10}/T_R\times (N_H\,k_{10}^{HH})/A_{10}$. After the inclusion of x-ray and VDKZ18 heating effects, the gas temperature remains $>10$~K. Therefore, we can take $k_{10}^{HH}\,\approx 3.1$ $\times10^{-11}$ $(T_{\rm gas}/{\rm K})^{0.357}$  $\exp(-32~{\rm K}/T_{\rm gas})$ ${\rm cm^3/sec}$ for ${\rm 10~K}<T_{\rm gas}<{\rm 10^3~K}$. As illustrated in plot \eqref{p_1} and \eqref{p_2}, increasing excess radiation fraction $A_r$ above $\sim 100$, the $T_R/T_S$ remains nearly constant and this also mean that $T_{21}$ remain unchanged. Consequently one can not increase the value of $B_0$ and one gets nearly flat profile for $B_0$ for $A_r\gtrsim100$ in figure \eqref{p_3b}. 


In figure \eqref{p_4}, we plot the maximally allowed values of $B_0$ vs $n_B$ for various values of $A_r$. The colour-bars represent the variation in $A_r$. In the plots, $A_r$ varies from 5 to LWA1 limit $\sim 418$. We consider both the EDGES best fit and upper constraint on 21 cm absorption signal for constraining $B_0$. The green-yellow scheme represent the case with  $T_{21}|_{z=17.2}\simeq-500$~mK, while the red-grey colour scheme represent the case  $T_{21}|_{z=17.2}\simeq-300$~mK. Numerical values of $A_r$ for the different colour bands are written in different colours. For $T_{21}|_{z=17.2}\simeq-300$~mK case the value of $A_r$ written with blue coloured text , while for $T_{21}|_{z=17.2}\simeq-500$~mK  case it is written with black coloured text. The spectral index ranges from -2.99 to -1.  The red dashed line represents the Planck 2015 upper constraint on the present-day magnetic field strength with spectral index in both plots. This constraint has been taken from Refs. \cite{Planck:2016, Minoda:2018gxj}.


In plot \eqref{p_4a}, we consider $T_{\rm S}\simeq T_{\rm gas}$ and we do not take into account the x-ray and VDKZ18 effects on IGM gas evolution. The zoomed inset in the figure shows the contour plot when $T_{21}|_{z=17.2}\simeq-300$~mK. Here, considering $T_{21}|_{z=17.2}\simeq-300$~mK,  for $n_B<-2.98$ the  $A_r\gtrsim 200$ is excluded similarly for $n_B<-2.96$ the $A_r\gtrsim 280$ is excluded by Planck 2015 upper constraint on $B_0$. Likewise, for $T_{21}|_{z=17.2}\simeq-500$~mK, for $n_B<-2.97$ the  $A_r\gtrsim 280$ is excluded. For spectral index -2.9 and excess radiation fraction 418, we get the upper constraint on $B_0$ to be $\sim 1$~nG and   $1.3$~nG by requiring $T_{21}|_{z=17.2}\simeq-500$~mK (EDGES best fit constraint) and -300~mK (EDGES upper constraint), respectively. While for $n_B=-1$,  these bound change to $1.1\times10^{-3}$~nG and $1.6\times10^{-3}$~nG for $T_{21}|_{z=17.2}\simeq-500$~mK and -300~mK, respectively. In plot \eqref{p_4b}, we include both the VDKZ18 and x-ray effect and consider finite Ly$\alpha$ coupling. As discusses earlier, for $A_r\gtrsim100$, $T_R/T_S$ ratio remain nearly constant. Therefore, in the plot \eqref{p_4b}, we can see that for $A_r\gtrsim100$, the upper bound on $B_0$ is not changing significantly---the plots are merged for $A_r\gtrsim100$. These plots have been shown by the zoomed inset. The right upper zoomed inset is shown for $T_{21}\simeq-300$~mK, while left lower zoomed inset is shown for green-yellow contour plots when $T_{21}\simeq-500$~mK. Therefore, further increasing $A_r>100$ will not change significantly the upper bound on $B_0$. As illustrated in figure \eqref{p_1} and \eqref{p_2}, $T_S> T_{\rm gas}$ for $A_r >0$, and $T_{21}\propto (1-T_R/T_S)$. Therefore, to get $T_{21}\simeq-300$~mK or $-500$~mK, we need to lower $B_0$ compared to previous scenario---figure \eqref{p_4a}. Hence, we get the more stringent upper bound on present-day magnetic field strength in figure \eqref{p_4b}. For spectral index -2.9 and excess radiation fraction 418, we get the upper constraint on $B_0$ to be $\lesssim 1.7\times10^{-1}$~nG and   $1.2\times10^{-1}$~nG by requiring $T_{21}|_{z=17.2}\simeq-300$~mK and -500~mK, respectively. For $n_B=-1$, we get $B_0\lesssim6.9\times10^{-5}$~nG and $3.7\times10^{-5}$~nG by requiring EDGES upper and best fit constraint on 21~cm differential brightness temperature. Decreasing the values of $A_r$, the upper constraint on $B_0$ becomes more stringent. For example, when $A_r=5$, we get upper bound on present day magnetic field strength to be $\lesssim1.4\times10^{-1}$~nG for spectral index -2.99, and for spectral index $n_B=-1$ we get $B_0\lesssim3.8\times10^{-6}$~nG by requiring EDGES best fit constraint on $T_{21}$. The upper bounds are also well below the Planck 2015 constraint \cite{Planck:2016}.


\section{Conclusions}
In the present work, we study the upper constraint on the strength of the primordial magnetic fields for different spectral index using the bound of EDGES observation on $T_{21} $, in the presence of uniform redshift-independent synchrotron like radiation reported by ARCADE 2 and LWA1 \cite{Fixsen2011, Feng2018, Dowell2018, Fialkov:2019}. We have considered excess radiation fraction up to the LWA1 limit at the reference frequency of 78~MHz, i.e. $A_r\sim418$  \cite{Dowell2018, Fialkov:2019}. To get the upper constraint on $B_0$, we have used both the EDGES upper and best-fit constraints on $T_{21}$.  We have considered two scenarios: First, infinite Ly$\alpha$ coupling (i.e. $x_\alpha\gg x_c, 1$) without the effects of x-ray and VDKZ18 on IGM gas evolution. Next, we have considered the finite  Ly$\alpha$ coupling with x-ray and VDKZ18 effects. The following summarises our results for $T_{21}=-500$~mK.

In the first scenario, for $A_r=418$, we get $B_0\lesssim3.7$~nG for spectral index -2.99, while for $n_B=-1$ we get $B_0\lesssim1.1\times10^{-3}$~nG. When $A_r=5$, upper constraint on present-day magnetic field strength varies from $B_0\lesssim 2.9\times10^{-1}$~nG to $1.8\times10^{-5}$~nG by varying $n_B$ from $-2.99$ to -1, respectively. 

In the second scenario, the upper bounds on $B_0$ will modify \cite{Venumadhav:2018, Kovetz2018}. For $A_r=418$, we get the upper constraint on magnetic field to be $B_0(n_B=-2.99)\lesssim4.9\times10^{-1}$~nG and $B_0(n_B=-1)\lesssim3.7\times10^{-5}$~nG. While for $A_r=5$, we get upper bound on present day magnetic field strength to be $\lesssim1.4\times10^{-1}$~nG for spectral index -2.99, and for spectral index $n_B=-1$ we get $B_0\lesssim3.8\times10^{-6}$~nG.

We would like to note that these upper bounds on $B_0$ that we have reported here are also consistent with the Planck observations \cite{Planck:2016, Planck:2014}.


\section*{Acknowledgements}
The author would like to thank Prof. Jitesh R. Bhatt for improving the presentation of the manuscript and helpful discussions. The author would also like to thank Prof. Karsten Jedamzik and Alekha C. Nayak for useful comments.  All the computations are accomplished on the Vikram-100 HPC cluster at Physical Research Laboratory, Ahmedabad. Finally, the author thanks the Referee for suggestions and a detailed report that significantly improved the quality of the manuscript.

%

\end{document}